\documentclass[sigconf]{acmart}
\usepackage{amsmath,amssymb,amsfonts}
\usepackage{algorithmic}
\usepackage{graphicx}
\usepackage{textcomp}
\usepackage{xcolor}
\usepackage{url}
\usepackage{hyperref}
\usepackage{setspace}
\usepackage{framed}

\AtBeginDocument{%
  }

\copyrightyear{2026}
\acmYear{2026}
\setcopyright{cc}
\setcctype{by}
\acmConference[ICSE-SEIP '26]{2026 IEEE/ACM 48th International Conference on Software Engineering}{April 12--18, 2026}{Rio de Janeiro, Brazil}
\acmBooktitle{2026 IEEE/ACM 48th International Conference on Software Engineering (ICSE-SEIP '26), April 12--18, 2026, Rio de Janeiro, Brazil}

\begin{document}

\title[Group Review Requests]{Group versus Individual Review Requests: Tradeoffs in Speed and Quality at Mozilla Firefox}

\author{Matej Kucera}
\affiliation{
 \institution{University of Groningen}
 \city{Groningen}
 \country{The Netherlands}}
\email{matej.kucera@yahoo.com}
\orcid{0009-0009-0343-7199}

\author{Marco Castelluccio}
\affiliation{
  \institution{Mozilla Corporations}
  \city{London}
  \country{UK}}
\email{mcastelluccio@mozilla.com}

\author{Daniel Feitosa}
\affiliation{
 \institution{University of Groningen}
 \city{Groningen}
 \country{The Netherlands}}
\email{d.feitosa@rug.nl}
\orcid{0000-0001-9371-232X}

\author{Ayushi Rastogi}
\affiliation{
 \institution{University of Groningen}
 \city{Groningen}
 \country{The Netherlands}}
\email{a.rastogi@rug.nl}
\orcid{0000-0002-0939-6887}

\begin{abstract}
The speed at which code changes are integrated into the software codebase, also referred to as code review velocity, is a prevalent industry metric for improved throughput and developer satisfaction.
While prior studies have explored factors influencing review velocity, the role of the review assignment process, particularly the `group review request', is unclear.
In group review requests, available on platforms like Phabricator, GitHub, and Bitbucket, a code change is assigned to a reviewer group, allowing any member to review it, unlike individual review assignments to specific reviewers. 
Drawing parallels with shared task queues in Management Sciences, this study examines the effects of group versus individual review requests on velocity and quality.

We investigate approximately 66,000 revisions in the Mozilla Firefox project, combining statistical modeling with practitioner views from a focus group discussion.
Our study associates group reviews with improved review quality, characterized by fewer regressions,  while having a negligible association with review velocity.
Additional perceived benefits include balanced work distribution and training opportunities for new reviewers. 
\end{abstract}



\maketitle

\section{Introduction} \label{sec:introduction}
Code review velocity is a key metric in the software development industry~\cite{mozilla2016, cd-facebook}, which also plays a significant role in developers’ job satisfaction~\cite{fb, mozilla2016, idle-time, nudge-fb, nudge-microsoft}. 
In open source software (OSS), maintaining high velocity is particularly crucial, as prolonged delays can discourage contributors~\cite{oss-abandon}. 
Moreover, long-lived patches can encounter problems such as communication hindrance and disconnection from the current state of the code, which can result in complex merge conflicts and integration issues~\cite{nudge-microsoft}.
However, prioritizing speed can come at the expense of other goals such as code quality~\cite{cd-facebook}, potentially undermining the review process. 
It is essential to monitor these trade-offs to ensure that the process remains effective while improving velocity.

There are many ways to expedite code reviews.
One way is to uncover factors influencing the review process, such as change size~\cite{MR-size} and author experience~\cite{mozilla2016, linux2013, microsoft2015}.
Building on these insights, there are solutions to reduce delays.
For example, reviewer recommendation systems to reduce the time to find a reviewer~\cite{whodo}, reminders for blocking reviewers~\cite{nudge-fb, nudge-microsoft}, and automated merges on acceptance~\cite{idle-time}. 
Solutions aiming to quantify factors influencing velocity in the literature often overlook the process of reviewer assignment, instead focusing on the review process itself and its outcomes, which occur after the reviewer has already been assigned. 

One reviewer assignment strategy is a group review request in which a code change is assigned to a group rather than an individual for review.
This feature is available on platforms such as GitHub, Phabricator, Bitbucket, and Azure DevOps~\cite{github-code-owners, mozilla-getting-reviews, bitbucket-reviewer-groups, azure-reviewer-groups}, allowing any team member to complete the review.
From anecdotal evidence, we observe that even when this feature is unavailable or not utilized, teams may resort to workarounds, such as posting open pull requests in group chats, suggesting its relevance. 

Studies in Queue Management research show that shared task queues (like group review requests) improve throughput~\cite{queues1976} by distributing work evenly and reducing delays by avoiding bottlenecks.
However, in practice, empirical studies suggest that individual queues (such as individual review requests) often perform better, due to a stronger sense of responsibility and fewer instances of social loafing~\cite{social-loafing, queues-ed, queues2018, queues-supermarket}, where no one steps up.

\begin{figure*}
    \centering
    \includegraphics[width=0.9\linewidth]{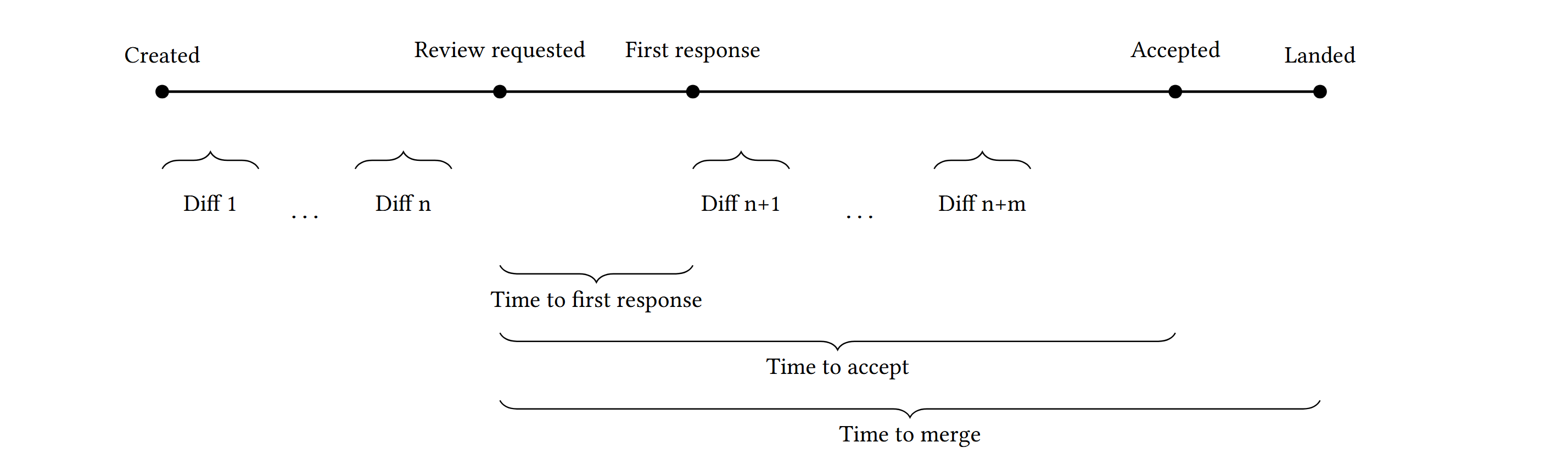}
    \caption{Life cycle of a revision}
    \Description{The picture shows the lifecycle of a Phabricator revision. A timeline shows the progress over time: creation, revisions before review, review request, first reviewer response, revisions, acceptance, landing. It also shows the span of time covered by time-to-first-response (from first review request to first reviewer response), time-to-accept (from first review request until last review acceptance) and time-to-merge (from first review request until landing). }
    \label{fig:revision_lifecycle}
\end{figure*}

This research aims to investigate how the review assignment process, whether group or individual, influences code reviews. 
We examine two dimensions, namely velocity and quality, to determine whether gains in speed come at the expense of quality or vice versa.
Therefore, our research question is: 

\vspace{1em}
\emph{How do group review requests influence code review velocity and quality compared to individual requests?}\\

We study Mozilla Foundation’s Firefox project, which employs both individual and group code reviews~\cite{mozilla-getting-reviews}. 
Firefox is a widely used open-source browser, serving millions of users daily~\cite{mozilla-user-activity-report}, with 240,226 revisions recorded on Phabricator as of March 2025 since its adoption in 2018~\cite{mozilla-phabricator-launch}.
To examine the impact of the reviewer assignment process, we built statistical models comparing the effects of individual versus group reviews on review velocity and quality. 
We complement this quantitative analysis with qualitative insights from experienced developers, gathered through a focus group discussion.

Our study associates group reviews with a lower likelihood of regressions, suggesting higher software quality compared to individual review requests.
These differences in software quality are observed alongside negligible to no difference in review velocity, as measured by time-to-accept.
Furthermore, developers believe that there are other advantages to group review as a review assignment process, including balanced workload distribution and opportunities to train new reviewers.

\section{Background and Related Work} \label{sec:background}

This section first outlines Mozilla’s code review process, then reviews existing literature on code review velocity and relevant Management Science studies on task queues.

\subsection{Code Review at Mozilla}
Mozilla uses a variety of tools to streamline its code review process.
It tracks bugs using its custom issue-tracking platform, Bugzilla~\cite{bugzilla}.
To solve a bug, an author submits a patch for review using the Phabricator code review platform.

Phabricator creates a \textit{revision} from a submitted diff.
A \textit{diff} is a set of changes submitted by the author to the version control system (VCS).
Authors can update a revision by submitting new diffs.
Phabricator’s stacked diffs feature allows large code changes to be split into multiple smaller revisions, which are generally easier to review.
The number of revisions in such a stack is referred to as the stack size~\cite{mozilla-stack-size}. 
Revisions are reviewed by other developers through manual or automated \textit{review requests}, which assign reviewers~\cite{mozilla-getting-reviews}.
Once a revision is accepted, it can be \textit{landed} (merged) into the codebase.

A typical code review at Mozilla, as illustrated in \autoref{fig:revision_lifecycle}, follows a structured sequence of events: first, a revision is authored and submitted, followed by the assignment or request of one or more reviewers.
These reviewers then offer feedback or approvals, after which the approved revision is integrated into the codebase.
Along the way, the process may involve intermediate steps such as updating the revision based on feedback, undergoing additional review iterations, and running automated checks or validations. 

Another common feature of Mozilla’s code review workflow is group review requests. 
Since June 2018, Mozilla has used Phabricator as its primary code review platform~\cite{mozilla-phabricator-launch}, which supports both manual and automated assignment of (group) review requests.
Authors can designate groups as reviewers for a revision, or have them assigned automatically through predefined `Herald rules,' which trigger based on revision characteristics (e.g., the revision contains changes to a specified file path).
Most of Mozilla’s Herald rules are configured to assign group reviews~\cite{herald-rules}.

The code review process in Phabricator generates a large amount of data.
Mozilla maintains \verb|bugbug|, an open-source platform for machine learning projects on software engineering tasks~\cite{mozilla-bugbug}, containing both ML models, tools, and datasets.
The bugbug code reviews dataset contains raw Phabricator logs, including revision transactions and metadata that describe the review process.

Given the two reviewer assignment processes — group and individual — it is crucial to understand their influence on the code review process. Is one approach more effective than the other? Or under what conditions does one approach outperform another?
Since both forms of review are widely practiced at Mozilla, this setting offers a valuable opportunity to address the question with empirical evidence.

\subsection{Code Review Velocity}
Code review velocity is influenced by many factors, as found by existing literature~\cite{whodo, mcr, webkit2013, microsoft2015, linux2013, MR-size, kernel-code, idle-time, nudge-microsoft, mozilla2009, google, nudge-fb, mr-latency}. There are four major entities these can belong to: author, reviewer(s), revision, and project. \autoref{tab:metrics} contains a list of factors investigated in previous studies, sorted by the entity they belong to. This extensive, yet non-exhaustive list was compiled through a literature search and analysis of a recent empirical overview~\cite{mr-latency2}. 

Several past studies have investigated code review at Mozilla, the latest of which was published in 2016~\cite{mozilla2002, mozilla2009, mozilla2015, mozilla2016}. 
Prior to the introduction of Phabricator, Mozilla used MozReview (a fork of ReviewBoard) as a code review platform. 
MozReview did not have support for group review requests~\cite{mozreview}. 
The process has undergone major changes since then with the introduction of Phabricator in 2018~\cite{mozilla-phabricator-launch}. 
To the best of our knowledge, no studies have been conducted on the code review process at Mozilla since the process change in 2018. 

\subsection{Task Queue Management}
Task queue management is the process of managing a queue of tasks to ensure they are completed efficiently, fairly, and in the right order~\cite{queues1976}. 
Task queue management is a complex problem, influenced by social and contextual factors, and comes in two configurations: a shared queue and an individual queue~\cite{queues-supermarket}. 
In a shared queue, a task is assigned to a pool of workers so that any member of the pool can pick a task. 
In an individual queue, a task is assigned to a single individual. 
In the context of code review, a group review request is equivalent to a shared queue and individual requests to individual queues.

\begin{table}[tb]
\caption{Factors known to influence code review velocity}
\label{tab:metrics}
\centering
\begin{tabular}{@{}ll@{}}
\textbf{Entity} & \textbf{Metric} \\
\midrule
Pull request & \textbf{Size in lines of code (LOC)}~\cite{linux2013, MR-size, nudge-microsoft, mr-latency} \\
 & \textbf{Modules affected}~\cite{mozilla2015, linux2013} \\
 & Area hotness~\cite{mr-latency, linux2013} \\
 & \textbf{File types}~\cite{nudge-microsoft} \\
 & \textbf{Number of reviewers}~\cite{linux2013, nudge-fb} \\
 & \textbf{Dead code removal}~\cite{mcr} \\
 & \textbf{Number of files}~\cite{mozilla2015, linux2013, nudge-fb, nudge-microsoft} \\
 & \textbf{Number of commits}~\cite{mr-latency, mr-latency2} \\
 & Has assigned reviewer~\cite{nudge-fb} \\
 & \textbf{Testing}~\cite{mr-latency, MR-size} \\
 & \textbf{Date}~\cite{linux2013, nudge-microsoft} \\
 & \textbf{Created Friday}~\cite{mr-latency, MR-size} \\
 & Description length~\cite{mr-latency, MR-size} \\
 & CI result~\cite{mr-latency, MR-size} \\
 & \textbf{Comments}~\cite{linux2013, MR-size, mr-latency, mozilla2015} \\
 & Priority~\cite{mozilla2015} \\
 & \textbf{Severity}~\cite{mozilla2015} \\
 & \textbf{Task type}~\cite{linux2013, nudge-microsoft} \\
\midrule
Project & CI latency~\cite{mr-latency, MR-size} \\
 & Project age~\cite{mr-latency, MR-size} \\
 & Team size~\cite{mr-latency} \\
 & Open revision count~\cite{mozilla2015, mr-latency, MR-size} \\
 & Mean patch latency~\cite{nudge-microsoft} \\
\midrule
Author & Relationship with reviewers~\cite{nudge-fb} \\
 & \textbf{Author affiliation}~\cite{mr-latency, MR-size, mozilla2016} \\
 & \textbf{Author seniority}~\cite{nudge-microsoft} \\
 & \textbf{Change experience}~\cite{linux2013, MR-size, mozilla2016} \\
 & \textbf{Review experience}~\cite{microsoft2015} \\
 & \textbf{Author success rate}~\cite{mr-latency} \\
\midrule
Reviewer & Reviewer workload \\
 & Reviewer experience~\cite{mozilla2015} \\
 & Blocking reviews~\cite{nudge-microsoft} \\
\bottomrule
\multicolumn{2}{l}{$^1$Metrics in \textbf{bold} are confounding factors}
\end{tabular}
\Description{Table containing a list of metrics from literature whihc have been investigated in relation to code review.}
\end{table}

The literature on task queue management presents two complementary views on how the shared task queue affects load balancing and efficiency.
One view is that with all else being equal, a shared task queue reduces labor cost without sacrificing latency~\cite{queues1976}.
The underlying idea here is that, given more potential workers, bottlenecks and excessive individual workloads can be mitigated.
However, this view is considered theoretical as it does not account for social, environmental, and other forces at play.
For example, shared task queues can exhibit the \textit{competition effect}, where workers compete with others sharing the queue for monetary, intangible, or other forms of compensation, if available~\cite{queues-supermarket}.
However, shared queues can reduce efficiency through \textit{social loafing}, where individuals deliberately reduce their efforts, anticipating that their colleagues will compensate for the delay.

Individual queues, in contrast, are associated with stronger task ownership and an increased risk of uneven load distribution and task abandonment. 
Studies show that workers can be motivated by a stronger sense of ownership when tasks are addressed to them~\cite{queues-ed}. 
However, when tasks are assigned following a pattern (e.g., seniority and visibility), the imbalance in queue length can cause bottlenecks and lead to delays. 

Code reviews at Mozilla utilize both group review requests and individual reviewer assignments; however, their impact on review velocity and quality remains unclear. 
Drawing on insights from task queue management, we observe multiple ways in which assignment strategies can influence review speed. 

In the specific context of Mozilla, we also aim to understand the reasons behind the adoption of each approach. 
Shared queues, for instance, may be useful when it is uncertain who is best suited to review a change, but it is not well understood how such queues function in a human-intensive, collaborative process like code review. 
To address these gaps, our study examines how review assignment practices impact both the quality and the velocity of code reviews.

\section{Study Design} \label{sec:method}
To gain a comprehensive understanding of how group review requests impact code review velocity and quality compared to individual review requests, we design a mixed-methods study.
First, we mine software archives to build statistical models, controlling for the effects of confounding factors, to compare the two review assignment processes in terms of velocity and quality.
Then, we present these insights to a curated group of experienced developers during a focus group to elicit their experiences around the observations.
Combined, the two approaches contribute to presenting a comprehensive view of the tradeoffs in velocity and quality depending on the choice of review assignment process.

The following subsections outline our design choices for analysis, beginning with the data processing approach used to answer our research question.
Next, we elaborate on our choice of metrics to measure review velocity, quality, and confounding factors.
Finally, we present details of statistical modeling and the focus group session that collectively answer our research question.
All source code and the dataset used in this study are available as a replication package on Figshare~\cite{replication-package}. 

\subsection{Dataset}
To analyze a typical code review workflow at Mozilla, we applied some inclusion and exclusion criteria. Refer to \autoref{sec:background} and \autoref{fig:revision_lifecycle} for a typical workflow.

\subsubsection{Inclusion and Exclusion criteria}
We study Mozilla's largest project, the Firefox browser, which has an extensive development history. 
The Firefox project's bug tracker, Bugzilla, is organized into 43 teams, where each team is responsible for one or more products, which in turn comprise one or more components. The teams (groups) are organized either by functionality (e.g., anti-tracking, cookies) or by project modules and components (e.g., Android, build).

We started with $240,226$ revisions from the Firefox project, which were further enriched with additional information from \verb|bugbug| (described in \autoref{sec:background}).
We selected all published revisions from the Firefox project created after the official adoption of Phabricator in June 2018~\cite{mozilla-phabricator-launch}.
The end cutoff date was the 8th of March 2025. 
We excluded nearly a year of data prior to official adoption, as it may not adhere to a typical code review workflow.

Next, we excluded revisions that were abandoned or in progress, since they had not yet completed the code review process. 
This should not influence the analysis as there was no correlation between review request type and revision status.
The subsequent exclusions were to improve data quality by removing anomalous and edge cases. 
Therefore, like prior works~\cite{MR-size}, we excluded revisions that skip the review process.
This includes revisions reviewed by the author or having no review.
We also removed revisions with inconsistent lifecycle timestamps (e.g., a revision marked as landed before it was created), resulting from authors manually publishing revisions without following the standard code review protocol.

Further, we excluded revisions that miss an associated bug report in Bugzilla. 
According to a senior researcher at Mozilla involved in the study design, an associated bug report is a sign of task relevance and likely adherence to a typical code review process, and vice versa. 
Finally, revisions with missing confounding attributes (details to follow), such as `issue type' and `diffs', were also excluded.
Our final dataset comprised $175,330$ revisions out of $240,226$ revisions, that is 70\% of all data. 

\subsubsection{Group requests masquerading as individual requests} \label{subsec:masquerading}
During a sanity check, we manually compared random code review data in Phabricator with its raw, extensive equivalent in the \verb|bugbug| dataset and information in the Phabricator UI, and observed an anomaly.
In some cases, we found that the reviewer deleted the group review request when accepting.
For example, refer to revision 189553\footnote{\url{https://phabricator.services.mozilla.com/D189553}}.
Later, we identify the reason for this behavior in focus groups; however, for the purpose of data analysis, these requests look like individual requests (since the group review request is deleted) and account for 3\% of all revisions.

To improve data quality, we supplement the review assignment process inferred from Phabricator with additional information from the \verb|bugbug| dataset, which enabled us to identify these deleted review requests.
The resulting data was more accurate, but also somewhat smaller, because \verb|bugbug| contains data for only the past two years. 
Our final dataset consisted of $66,318$ revisions.
We also ran our experiments on the larger dataset, which contains 173,701 revisions, for comparison, and observed similar results (details in replication package~\cite{replication-package}). 

\subsection{Metrics}
To study the influence of group versus individual review requests on review velocity and quality, we select their closest proxies from the literature that can be measured using the code review data available for the Firefox project. Likewise, we identify measurable confounding factors from the literature and those specific to the project context. 

\subsubsection{Code review velocity}
In a code review lifecycle, there are three key events: first-response, accept, and merge, which indicate the review's velocity. 
Likewise, there are three commonly used metrics for measuring code review velocity.
Time-to-first-response measures the time until the initial response by a reviewer~\cite{idle-time}.
Time-to-accept is the time until a revision is accepted by (all) reviewers~\cite{idle-time}.
Finally, time-to-merge refers to the time until the revision is merged into the codebase~\cite{idle-time}. 
The relationship between the three metrics and their span over the revision life cycle is graphically depicted in \autoref{fig:revision_lifecycle}.

We choose time-to-accept as an indicator of review velocity, as it best captures the part of the review cycle likely influenced by group review requests. 
In contrast, time-to-merge includes other causes of delay related to merging, which are not influenced by reviewers.
Time-to-first-response could have been another plausible candidate, as it captures the first activity after individual versus group assignment, which is only a part of the whole. 
We chose not to use it because it does not show the whole picture, and time-to-accept is more relevant for Mozilla. 

We measure time-to-accept as the time between the creation of the first review request and revision acceptance~\cite{tta}. 
Because Phabricator allows authors to publish a draft revision and modify it before requesting a review (see \autoref{fig:revision_lifecycle}), we modify the definition of time-to-accept and define the start point as the creation of the first review request rather than the revision itself.
In our data, 5\% of all revisions are 8 hours or more apart in terms of time of creation and first review request. 

\subsubsection{Group review requests}
As previously explained, there are two types of review requests: group requests and individual requests.
However, a single revision can be associated with multiple review requests, created asynchronously. 
We classify revisions based on their initial review request(s) as: group-only, mixed, and individual-only. 
These classes account for 33.2\%, 4.2\%, and 62.7\% of revisions, respectively.

\subsubsection{Code review quality}
Assessing the quality of code reviews is difficult as reviews aim to fulfill multiple, and often intangible, objectives. 
One metric considered important at Mozilla is the number of regressions, bugs that bypass the review process, since these directly affect the end-user experience. 
Regressions are defined as errors introduced by changes that disrupt previously functioning features. 
Mozilla systematically monitors known regressions~\cite{mozilla-regressions}.

In our dataset, we annotated each revision according to whether it was subsequently identified as causing a regression after landing. 
This measure was adopted as a proxy for code review quality. 
Within our filtered dataset, 25.6\% of revisions were identified as causing regressions. 
However, it should be noted that regressions may remain undiscovered for recent revisions, which limits the completeness of this metric.
This should affect both individual and group review requests in the same way; however, it is a potential source of bias.

\subsubsection{Confounding factors}
We control for factors known to influence code reviews, as well as factors specific to Mozilla. 
Referring to the literature on the topic (see \autoref{tab:metrics}), we include all factors measurable in the dataset (highlighted in bold). 
Furthermore, we include factors specific to Mozilla (see \autoref{tab:specific-metrics}) as studies have shown that context, such as project age and team size~\cite{overview2022}, influences review velocity. 
Finally, we gather unique features of the Phabricator and Bugzilla ecosystems, inspired by Phabricator and Mozilla documentation~\cite{mozilla-stack-size} and experiences of a senior Mozilla engineering manager. 

\begin{table}[tbp]
\caption{Context-specific metrics}
\label{tab:specific-metrics}
\centering
\begin{tabular}{@{}ll@{}}
\textbf{Entity} & \textbf{Metric} \\
\midrule
Revision & \textbf{First review request type}\\
 & \textbf{Stack size}~\cite{mozilla-stack-size} \\
 & \textbf{Created on weekend} \\
 & \textbf{Causes regression} \\
 & \textbf{Fixes regression} \\
 & \textbf{Team} \\
 & \textbf{Number of comments on issue} \\
\midrule
Author & \textbf{Author affiliation} (staff, contributor) \\
\bottomrule
\end{tabular}
\end{table}

\subsection{Statistical Modelling}
We experimented with various preprocessing steps to build a representative statistical model. 
We noticed that the distribution of time-to-accept and LOC changed is highly skewed; therefore, we experimented with three thresholds, namely, 95\%, 99\%, and none~\cite{statistics} for these fields. 
Furthermore, in 2020, Bugzilla revised its severity measurement method from descriptive (e.g., `critical`, `major`) to coded (e.g., `S1`, `S4`). 
Therefore, we built models with valid severity entries only, or with the severity field removed. 
Finally, we experimented with deleted review requests (explained in \autoref{subsec:masquerading}) by including or excluding them. 
This resulted in 36 datasets ($3\cdot3\cdot2\cdot2$) ranging from $175,459$ to $23,186$ revisions (more detail in replication package~\cite{replication-package}). 

Next, we build multiple linear regression models to study the influence of review request type on review velocity. 
The choice of a linear regression model enables us to find the relationship between review request type and review velocity while controlling for the effects of confounding factors. 
We consider a relationship statistically significant if the p-value $<0.05$.
We log-transformed skewed features, such as LOC added and removed, for standardization and to stabilize variance and reduce heteroscedasticity~\cite{linreg}. 

We also assessed features for multicollinearity using the Variance Inflation Factor (VIF)~\cite{vif}. 
All VIFs were below 3, suggesting no multicollinearity problems.
Finally, we one-hot encoded all categorical variables, representing them as binary vectors of 0 and 1, as is required by a linear regression.

We assess the model fit using adjusted $R^2$~\cite{linreg} and use Gr\"{o}mping's method~\cite{relimp} to measure the relative importance of each feature. 
This method measures effect size by calculating the decrease in $R^2$ when removing each variable, thereby determining the percentage of the model explained by each variable.

Likewise, to study the influence of review type on review quality, we build a logistic regression model~\cite{logreg}. 
A logistic regression has the same benefits as linear regression with a Boolean dependent variable. 
We assess the model fit using McFadden's Pseudo-R$^2$~\cite{mcfadden-rsquared}. 
We do not calculate effect size because there is no established method for calculating it in logistic regression models.

\subsection{Focus group}
We conducted a presentation and focus group discussion with Mozilla developers to complement the quantitative analysis. 
Statistical results alone risk overlooking contextual factors, particularly in industrial settings where organizational processes, cultural practices, and tool-specific behaviors shape outcomes. 
Engaging practitioners enabled us to validate our findings, align our conclusions with real-world practices, and incorporate interpretations grounded in their experience and expertise. 
Participants were invited directly by a senior engineering manager through internal channels and via a broadcast invitation. 
The invited participants were selected based on their reviewer expertise and/or interest in the code review process.
Participation was voluntary, and the session was recorded with informed consent.

The session consisted of two components: a 15-minute presentation of results followed by a focus group discussion~\cite{focusgroup}. 
The presentation covered outcomes from the statistical analyses, including the linear regression model on code review velocity, the logistic regression model on review quality, and correlation checks for alternative explanations.
To broaden participation, we also distributed a short survey for those unable to join the discussion. 
Data from the session (recording and transcript) and the survey responses were analyzed for themes~\cite{thematic-analysis}.

\begin{table*}[tbp]
\centering
\caption{Linear regression model inferring code review velocity (coefficients in seconds)}
\begin{tabular}{l|r|r}
\textbf{Variable} & \textbf{Coefficient (error)} & \textbf{Relative Importance} \\
\midrule
Intercept & 314329 (19178)*** & N/A\\
num\_diffs & 267091 (2955)*** & 46.72\%\\
number\_nonauthor\_comments & 65907 (3453)*** & 17.11\% \\
number\_author\_comments & 40203 (3387)*** & 15.44\%\\
log(loc\_source\_code\_added) + 0.5 & 57230 (3587)*** & 4.73\%\\
stack\_size & 75001 (2616)*** & 4.30\% \\
number\_of\_reviewers & 34880 (2413)*** & 4.18\%\\
causes\_regression & 29509 (5406)*** & 1.47\%\\
issue\_type\_ENHANCEMENT & 30616 (6948)*** & 1.06\% \\
author\_success\_rate & -7366 (2446)** & 0.89\% \\
fixes\_regression & -58744 (7444)*** & 0.85\%\\
previous\_accepted\_revisions & -15240 (2511)*** & 0.84\%\\
author\_past\_review\_fraction & -23449 (2807)*** & 0.58\% \\
log(loc\_source\_code\_removed) + 0.5 & -16691 (2994)*** & 0.58\% \\
is\_staff & -23428 (5774)*** & 0.41\% \\
created\_friday & 52821 (5842)*** & 0.23\%\\
contains\_tests & -22026 (5525)*** & 0.20\%\\
deletions\_only & 36577 (13337)** & 0.18\% \\
issue\_type\_TASK & -13278 (5752)* & 0.08\% \\
first\_reviews\_are\_group\_MIXED & -4973 (11447) &  0.06\% \\
\textbf{first\_reviews\_are\_group\_GROUP-ONLY} & \textbf{-12533 (5345)*} & \textbf{0.04}\% \\
files\_modified & -5720 (2223)* & 0.03\% \\
created\_weekend & 39228 (9711)*** & 0.02\%\\
author\_past\_changes\_fraction & 2907 (2917) & N/A \\
is\_reformat & -475 (51526) & N/A \\
issue\_comment\_count & 673 (2569) & N/A \\
\hline
\multicolumn{3}{l}{***p $< 0.001$, **p $< 0.01$, *p $< 0.05$}
\end{tabular}
\label{tab:regression_coefs}
\Description{The coefficients and relative importance for all metrics computed by linear regression.}
\end{table*}

\section{Results} \label{sec:results} 

\begin{table*}[htbp]
\centering
\caption{Logistic regression model inferring regression as code quality indicator}
\begin{tabular}{l|r|r}
\hline
\textbf{Variable} & \textbf{Coefficient (error)} & \textbf{Odds Ratio} \\
\midrule
Intercept & -1.37 (0.08)*** & 0.25 \\
stack\_size & 0.50 (0.01)*** & 1.65 \\
issue\_type\_ENHANCEMENT & 0.40 (0.03)*** & 1.49 \\
\textbf{first\_review\_type\_GROUP-ONLY} & \textbf{-0.36 (0.02)***} & \textbf{0.70} \\
deletions\_only & -0.26 (0.07)*** & 0.77 \\
created\_weekend & -0.21 (0.04)*** & 0.81 \\
log(loc\_source\_code\_added) + 0.5 & 0.21 (0.02)*** & 1.23 \\
is\_staff & 0.19 (0.03)*** & 1.21 \\
num\_diffs & 0.18 (0.01)*** & 1.20 \\
fixes\_regression & -0.17 (0.03)*** & 0.84 \\
created\_friday & -0.12 (0.03)*** & 0.89 \\
is\_reformat & -0.11 (0.22) & 0.90 \\
log(loc\_source\_code\_removed) + 0.5 & 0.09 (0.01)*** & 1.10 \\
author\_past\_review\_fraction & 0.07 (0.01)*** & 1.07 \\
time\_to\_accept\_seconds & 0.06 (0.01)*** & 1.06 \\
author\_past\_changes\_fraction & -0.06 (0.01)*** & 0.94 \\
previous\_accepted\_revisions & 0.06 (0.01)*** & 1.06 \\
first\_review\_type\_MIXED & -0.06 (0.05) & 0.94 \\
issue\_type\_TASK & 0.05 (0.03) & 1.05 \\
number\_author\_comments & 0.04 (0.01)** & 1.04 \\
contains\_tests & -0.04 (0.02) & 0.96 \\
files\_modified & -0.01 (0.01) & 0.99 \\
number\_of\_reviewers & 0.01 (0.01) & 1.01 \\
author\_success\_rate & 0.00 (0.01) & 1.00 \\
number\_nonauthor\_comments & 0.00 (0.01) & 1.00 \\
\hline
\multicolumn{3}{l}{***p $< 0.001$, **p $< 0.01$}
\end{tabular}
\label{tab:logistic_regression_coefficients}
\Description{The coefficients and odds ratios for all metrics computed by logistic regression.}
\end{table*}

Our models exhibited a wide range of performance, with adjusted $R^2$ values varying from $0.09$ to $0.33$.
In the context of linear regression, this constitutes a weak-to-moderate fit ($R^2$ values between $0.25$ and $0.50$~\cite{rsquared}).
We discuss this in \autoref{sec:threats}.
Our best performing model (with an adjusted $R^2$ of $0.33$) excluded the \textit{severity} feature, excluded revisions with deleted reviews, and excluded revisions with time-to-accept above the 99th percentile. As mentioned previously, this dataset contained $66,318$ revisions. 
Our second-best-performing model included deleted reviews, resulting in a larger dataset of $173,701$ revisions and an adjusted $R^2$ value of $0.31$. Due to reasons outlined in \autoref{sec:method} regarding deleted reviews, all subsequent analyses are based on our best-performing model. 

\subsection{Influence on Code Review Velocity}
\autoref{tab:regression_coefs} presents a multiple linear regression to explain time-to-accept (in seconds) as a function of the review assignment process.
The features are ordered by relative importance (effect size).
Their relationship is mediated by a list of confounding factors inferred from the literature, and is otherwise specific to the project.
We create a visual distinction of one-hot encoded categorical variables by representing them in uppercase.
For example, the `enhancement' \textit{issue\_type} is referred to as
`issue\_type\_ENHANCEMENT'. 
For brevity, we excluded team names from \autoref{tab:regression_coefs} despite their significance in 17 out of 33 cases.
The complete model is available in the replication package~\cite{replication-package}.

This model has an adjusted R$^2$ ($0.33$), which is comparable to other similar models in the literature~\cite{mr-latency}. 
Generally, a relatively low goodness of fit is expected, as code review is a complex socio-technical process, and many factors, such as private communications, cannot be fully accounted for. 
Finally, the Relative Importance column in \autoref{tab:regression_coefs} reports the effect size, with some exceptions. 
We do not report the relative importance of \textit{author\_past\_changes\_ratio}, \textit{is\_reformat}, \textit{issue\_comment\_count}, and \textit{team\_name} due to algorithmic runtime complexity in the order of $O(2^n)$. 
Note that we do not report the relative importance of non-statistically significant features (see \autoref{tab:regression_coefs}).

The results show that 79\% of the model is explained by the number of diffs and the count of author and non-author comments.
The size of revision, measured as the LOC added and stack size, explains another 9\% of the model. Similar to previous works~\cite{nudge-microsoft}, the reviewer count is important, with 4\% relative importance. 
All other factors, including the review assignment process, have a relative importance of less than $1.5\%$. 
Regarding the review assignment process, we observe that group review requests, measured as `first\_reviews\_are\_group\_GROUP-ONLY', are associated with a negligibly reduced time-to-acceptance (effect size: $0.04\%$) compared to individual requests.
\vspace{1em}

\begin{framed}
\noindent \textbf{Group and individual review requests are associated with similar code review velocity. }
\end{framed}

\subsection{Influence on Code Review Quality}
The model for inferring code review quality achieved a McFadden's Pseudo-R$^2$ of $0.10$~\cite{mcfadden-rsquared}. 
This is a weak-to-modest model fit~\cite{mcfadden-rsquared, pseudo-r2} and for the same reason: code review is a complex socio-technical problem.
\autoref{tab:logistic_regression_coefficients} presents the model coefficients and odds ratios, indicative of relative relevance. 
An odds ratio of 2 means that a unit increase in the predictor variable doubles the odds of the outcome.
In \autoref{tab:logistic_regression_coefficients}, features are ordered by the absolute value of the coefficient. 
Here as well, we removed team names from \autoref{tab:logistic_regression_coefficients} for ease of reading (see details in replication package~\cite{replication-package}).

The model shows that \textit{deletions\_only}, first review type, issue type, stack size, and LOC added are the factors most strongly associated with code review quality, measured in terms of regressions.
Some of these relations are expected.
For example, enhancements often cause more regressions than defects, or deletion-only revisions result in fewer regressions.
Likewise, LOC added and the stack size have a positive coefficient, suggesting that large code changes are more likely to cause regressions.

Most importantly, we find that `group-only' review requests are associated with a lower likelihood of regression compared to `individual-only' review requests with statistical significance ($p < 0.001$) and an odds ratio of $0.7$.
This insight assumes that developers opt for group and individual review requests for the same types of code changes, negating any underlying bias in the choice of review assignment process.
We explore this claim further below, where we seek plausible alternative explanations for the observation in the data and later through developers' experiences.

\subsubsection{Alternate explanations} 
\label{subsec:altex}
To further explore the relationship between group reviews and regressions, we seek other factors that may mediate this relationship. 
We suspect the reason may be the differences in how group reviews work (e.g., increased reviewer participation), as well as individual and situational factors (e.g., experienced developers tend to prefer group reviews).
We explore four such relations below, two of which are confounding factors from \autoref{tab:metrics} and \autoref{tab:specific-metrics} that have correlations with group versus individual requests and have a (relatively) high effect size. 

Our data is not normally distributed; therefore, we choose non-parametric tests.
For continuous and categorical variable pairs, we use the Mann-Whitney U test~\cite{mann-whitney} with Bonferroni correction~\cite{bonferroni} to measure statistical significance. 
Further, we measure their effects using Cliff's Delta (CD), with values less than $0.147$ suggesting a negligible effect~\cite{cliff-delta}. 
For categorical variable pairs, we measure significance using the Chi-squared test~\cite{chi-squared} and measure effect using Cramér’s V (CV), with values below $0.10$ suggesting no effect~\cite{cramer-v}.

\vspace{1em}

\emph{Reviewer Count}
One possibility is that group review requests solicit more reviewers than one, which as per Linus' law `given enough eyeballs all bugs are shallow'~\cite{raymond1999cathedral} would imply that group review requests garner comprehensive attention resulting in fewer regressions.
However, our results show no statistical difference (p-value = $1$) with an average number of completed reviews of $1.16$ for both group and individual review requests. 

Although there are no differences in the number of reviewers, the mechanism for selecting a reviewer varies. 
In case of individual review requests, the reviewer is selected by the code author. 
In contrast, in group review requests, one or more developers, who are members of the group, decide who should review the code.
This mechanism of selecting an expert reviewer can play a mediating role in reducing regressions within group review requests. 

\vspace{1em}
\emph{Herald rules}
Our previous exploration suggests that Herald rules mediate the distribution of review requests to individuals versus group(s). 
Therefore, we investigate differences between reviews requested by Herald and reviews requested by revision authors. 
In our data, 18.6\% of reviews are requested by Herald rules.

Herald rules are invoked irrespective of whether the code author knows who to ask for reviews or not. 
Our analysis shows that these rules favor group review requests heavily (CV $= 0.513$).
Additionally, the adoption of Herald rules is associated with increased discussion and the selection of issue type. 
We observe that assignments with Herald rules receive more comments from authors (CD = $0.091$) and reviewers (CD = $0.079$), although the relationship is negligible with the choice of reviewer request mechanism, also known as the first review type.
So, more discussions in group requests do not explain the observation.

Likewise, we observe that Herald rules are used more often to assign reviews for issue-type tasks than defects (CV $= 0.084$). 
However, the effect size between the first review type and issue type is negligible (CV = $0.052$), suggesting the type of issue does not explain why group review requests have fewer regressions. 

Next, we examine code author characteristics that can influence the selection of the first review type. 

\vspace{1em}
\emph{Author affiliation}
Mozilla documentation recommends that first-time contributors request individual reviews from their mentors~\cite{mozilla-how-to-contribute}. 
This suggests that a pattern may exist in who works on individual versus group review requests.
Upon further exploration, we find that staff are more likely to request group reviews (CV = 0.076).
Additionally, the choice of tasks varies, with contributors being more likely to work on enhancements than staff (CV = 0.092).
Furthermore, experienced developers at Mozilla note that contributors tend to prefer working on exciting features rather than addressing defects. 

Therefore, we investigate the relationship between author affiliation (staff or non-staff) and code regression, assuming author affiliation mediates the relation between review assignment approach and regression.
However, we did not find any relationship (negligible effect size: CV $= 0.021$), suggesting that author affiliation is not the reason for the fewer regressions in group review requests.

\vspace{1em}
\emph{Author expertise}
We believe that affiliation did not explain the relation since the author's expertise, irrespective of affiliation, could be more important. 
Since non-staff can be experts and non-experts, as also seen in the negligible effect size in the relationship between affiliation and expertise (CD$= -0.019$), we rather look at author expertise. 

We found that experts, measured by prior revision success rate (percentage of accepted revisions), tend to make more individual-only requests (CD$= 0.102$).
However, the expertise is found to be unrelated to either velocity or quality. 
Further inspection also showed that the ratio of accepted and abandoned revisions for groups and individuals is not different (p-value = 0.191). 

This relation does not appear to be influenced by developer traits (affiliation and expertise), code characteristics (task versus defect), reviewer assignment method (automated versus manual), or review process factors (reviewer count and discussion length).
It may relate to reviewer choice, as self-selected reviewers, in the case of group review requests, appear more effective than author-selected ones; however, further investigation of this topic lies outside the scope of the study.
Another possible explanation is that some teams that rarely use group reviews may be associated with more regressions and vice versa.

\begin{framed}
\noindent \textbf{Group review requests are associated with fewer regressions and, consequently, higher code quality compared to individual requests.}
\end{framed}

\subsection{Focus Group and Survey} \label{subsec:focus_group_results}
We conducted a focus group session on May 23, 2025, to complement our understanding from quantitative data analysis with developer experiences. 
The focus group was preceded by a 15-minute presentation summarizing the work and followed by targeted questions asked in person and via survey for those who could not stay beyond 15 minutes. 
The questions asked in the survey are available in the replication package~\cite{replication-package}.
We identified and invited nine senior Mozilla developers (with 10+ years of experience) working in various teams in Firefox, globally. 
At the end of the presentation, five members stayed for an additional 45 minutes, while the remaining four answered questions via a survey.  

Our questions to developers followed five key themes. 
For orientation and warm-up, we solicited developer experiences surrounding the purpose of group code reviews and the reason for deleting group review requests in the data.
Here, the question about the purpose of group code reviews sought to gather reasons not previously known, while the latter aimed to understand whether the discrepancy observed in the data was intentional or an error. 
All subsequent questions solicit their experience and understanding of factors influencing code review velocity and code quality in relation to group versus individual review requests. 

\subsubsection{Purpose of group review requests}
Previously, we observed that staff request more group reviews, while newcomers are advised to request reviews from their mentors.
Developers complemented our understanding by adding that group review requests are also used when an author is unsure of who to ask for review.
For example, a newly hired team member (also a staff member) can request a group review when unsure who should review a revision, as it takes time to build knowledge of which reviewers are best suited for different parts of the codebase. 

Furthermore, several developers note that group review requests are preferred for load balancing purposes. 
They used phrases such as ``because it is spreading the load'' and ``in order to load-balance'' to express the relevance of group review requests. 
Developers argue that, in most cases, group review requests spread the load and that if group review requests were not used, reviews would be assigned to the same few knowledgeable developers, thus overloading them while also blocking new team members from building knowledge. 
As a nuance, they add that this model does not prevent experts from examining critical changes, since blocking group review (most Herald-requested reviews are blocked~\cite {herald-rules}) is often viewed by more people. 
Additionally, they note that sometimes only one reviewer is qualified to respond, which limits the benefits of a group review request.

\subsubsection{Reason to delete group review requests}
During data checks, we observed that a significant number of group review requests were eventually deleted (details in \autoref{sec:method}). 
This varied from $13.9\%$ to $0.6\%$ across teams, accounting for $3\%$ of the entire dataset. 
Our attempts to improve data accuracy, however, reduced its size. 
We asked developers whether the deletion was in error or if there was a reason. 
In response, developers noted that the practices around group review requests vary across teams and over time. 
For instance, some teams delete group review requests after a member picks them for review to avoid other members spending time on them. 
Other teams, however, do not make any edits to ensure that other team members can also see it. 

\subsubsection{Factors influencing code review velocity}
We solicited developers' views on factors influencing review velocity and perceived actionable steps. 
In general, developers' perceptions of factors influencing code review velocity align with existing data and prior studies. 
Participants suggest that LOC changed, length of discussion, author experience, reviewer workload, and author-reviewer relationship influence review speed. 
When asked specifically as a reviewer, developers suggest that a clear summary, small changes, a few modified files, familiarity with the code, and trust in the author make reviewing easier and faster.
They further add that reviews that frequently alternate between requiring responses from the author and the reviewer progress more quickly than those stalled in one state due to missed questions or updates.

\subsubsection{Group review requests and fewer regressions}
Finally, to explore potential explanations for the relationship between group review requests and improved code quality, measured as fewer regressions, we investigated how factors such as author affiliation and the use of Herald rules can influence it. 
Note that the participants of the focus group were all staff members.
Therefore, their views present a partial picture.
Although given their role as mentors and reviewers for non-staff contributors, they may have some understanding. 

\vspace{1em}
\emph{Author affiliation}
Participants agree with the recommendations in the documentation, suggesting that newcomers request reviews only from their mentors. 
They further shared nuances that can influence the observed regression counts. 
Supporting the observation that patches from staff comprise more regressions (as seen in \autoref{tab:logistic_regression_coefficients}), they add that staff typically work on relatively complex tasks and have tight deadlines.
Furthermore, developers review patches from non-staff members with greater rigor to ensure higher quality standards compared to those from staff. 
A developer adds that ``you don't know if contributors are going to stick around or not'', so with a thorough review, the aim is to prevent future regressions. 
Another survey response reinforced the prior observation by referring to staff as `they will fix regressions quickly' while the same cannot be said for non-staff. 
Yet another view was that non-staff contributors may prioritize avoiding regressions more than staff to be seen as a good contributor. 

\vspace{1em}
\emph{Herald rules}
Finally, we asked participants about their experiences with Herald rules, since these are strongly associated with group review requests.
All participants reported that their teams use Herald.
As reviewers, they generally do not notice whether a review was requested automatically by Herald or manually, and they perceive no difference in the process.
Several developers explained that their teams configure Herald to enforce blocking group review requests.
Others added that they depend on Herald when they are unsure who should review certain code changes, noting that `Herald rules help new people get a review'. 

\begin{framed}
\noindent \textbf{Practitioners note that while review assignment practices vary across teams, group review requests help in load balancing, reviewer skill development, and broader visibility. }
\end{framed}

Additionally, developers generally agree with the models that explain code review velocity and quality, and often provide additional details and explanations for unexpected relationships (e.g., staff code changes resulting in more regressions).

\section{Discussion}
\label{sec:discussion} 
\subsection{Interpretation of results}
At the onset of this research, two contrasting theories from management sciences emerged, suggesting that group review requests can increase review velocity by avoiding delays and bottlenecks, or alternatively, decrease it with a reduced sense of responsibility and social loafing.
Our study suggests that the review velocity of group review requests is comparable to that of individual requests. 
This contradicts our expectations but can point to various plausible explanations. 
Perhaps both forces are at play, balancing out each other's influence. 
Alternatively, extensive delays in individual requests balance out the observation in group review requests and vice versa. 
While velocity remained comparable, our subsequent explorations on code quality and developers' views on the purpose of group review requests painted a comprehensive picture. 

Our study shows that group review requests are linked with a significant decrease in the chances of regression, a crucial indicator of review quality at Mozilla and Firefox, in particular. 
Furthermore, we examined several plausible factors that may mediate the relationship, including the experience and affiliation of developers associated with the use of group review requests, the scenarios in which group requests are preferred, and the differences in the workings of group versus individual review requests. 
None of the above factors explained why group review requests have fewer regressions. 

The only plausible explanation is the difference in the choice of reviewers in the two approaches. 
Individual requests assume that the code author knows a good candidate for review, 
while group review requests rely on self-selection based on interest, availability, and expertise.  
More research would be required to substantiate this relation. 

Additionally, developers participating in the focus group emphasize the relevance and preference for group review requests to identify potential reviewers when it is unclear, balancing workload, reviewer skill development, and broader visibility.  

In a nutshell, group review requests have comparable code review velocity to individual requests, but are linked with improved code quality via reduced regressions. 
Other perceived benefits include improved management of review tasks and the development of necessary skills for reviewing.

\subsection{Recommendations for industry}

We suggest organizations consider using group review requests, in addition to individual requests, especially where the focus is efficient management of reviews and quality improvements. 
We believe that insights from our research would also apply to similarly large projects where teams are organized in a hierarchical structure. 
The hierarchy enables assigning a patch for review to a group where developers most likely have the right expertise and interest.  

\subsection{Recommendations for academia}
From a research perspective, many questions remain unanswered. 
First, \emph{what is the differential impact of group versus individual review requests on other key code review objectives, such as usefulness, knowledge transfer, and long-term code maintainability?}~\cite{microsoft2015}.
Such research would contribute to presenting a comprehensive view of the relevance and limits of group review requests. 
For instance, our study has not factored in the costs of group review requests, which might improve review quality at the cost of additional time spent by developers to identify who should review.

Another possibility is to \emph{study the evolution of review skills and review velocity with time.} 
With group reviews encouraging more people to review, review velocity may increase. 
Or perhaps reviewers who gain experience through group reviews will also become faster with individual reviews.
An interesting variant would be a comparison of the impact on velocity over time for different teams, where some teams stay with individual reviews and others switch to group reviews.

\emph{How does the distribution of review load vary in group versus individual review requests?} 
In organizations like Mozilla, a few reviewers handle a disproportionate number of revisions, creating process bottlenecks.
Inspired by the developer experiences in our focus group, it would be interesting to explore whether and how group review requests demonstrably alleviate the pressure on these key individuals.

Finally, it will be interesting to dig deeper into \emph{why group reviews are associated with fewer regressions and which types of regressions are caught in group versus individual reviews}.

\section{Threats to Validity} \label{sec:threats} 
This research is prone to various threats to validity, as discussed below. 
We also present the choices we made to minimize the risks. 

\emph{Construct Validity}
Our models use proxies to approximate real-world characteristics. 
For instance, we measure author experience using `past changes' as a proxy. 
To minimize the risks associated with using a single proxy and its limitations, we employ multiple proxies to gather the same feature. 
Some other proxies used to measure author experience include previously accepted revisions, success rate, and reviews. 

At times, we were limited by the unavailability of more accurate measurement proxies.
For instance, prior works show that the number of LOC added and removed is less accurate, unless the number of lines changed is considered~\cite{MR-size}.
Unfortunately, the available dataset did not contain this information.  
Likewise, using `first review request' as a proxy to infer group review requests has its limits. 
For instance, information on the reassignments of the reviewer is not available. 

Using regressions as a proxy for code review quality has its limits. 
Since regressions are identified retrospectively, it is possible that some regressions may not have been discovered yet, especially for recent revisions.

\emph{Internal validity}
The validity of our findings may be limited because the models used in this study do not fully capture the underlying factors driving changes in the dependent variable. 
The multiple regression model explains only about one-third of the variance with an R$^2$ value of 0.33, while the logistic regression model shows similarly low explanatory power (McFadden’s Pseudo-R$^2$ of $0.10$.
Nevertheless, these values are consistent with similar prior studies~\cite{mr-latency}.

\emph{Generalizability}
This study relies on data from a single source, the Firefox project, presenting a threat to external validity. 
We resorted to this approach since the use of group review requests was scarce in projects other than Firefox. 
As a result, the findings may not fully generalize to other projects or organizations and should be validated in additional contexts.

\vspace{-0.1em}
\section{Conclusions} 
In this study, we examined over 66,000 revisions from Mozilla’s Firefox project to assess how group review requests compare with individual requests in shaping the code review process. 
Using statistical modeling and focus group discussions, we found that group review requests are associated with higher review quality without noticeable differences in velocity.
Additional advantages include better workload distribution, opportunities for reviewer growth, and increased visibility. 
Overall, the findings suggest that group review requests offer an effective means of enhancing the review process and outcomes while maintaining velocity.

\balance
\bibliographystyle{ACM-Reference-Format}
\bibliography{bib}

\end{document}